\documentclass[journal,twocolumn,10pt,twoside]{IEEEtran}

\normalsize

\usepackage{cite}

\ifCLASSOPTIONcompsoc
    \usepackage[caption=false, font=normalsize, labelfont=sf, textfont=sf, subrefformat=parens,labelformat=parens]{subfig}
\else
\usepackage[caption=false, font=footnotesize]{subfig}
\fi

\ifCLASSINFOpdf
  \usepackage[pdftex]{graphicx}
\else
\fi

\usepackage[flushleft]{threeparttable}
\usepackage[cmex10]{amsmath}
\usepackage{xcolor}
\usepackage{xfrac}
\usepackage{lipsum}
\usepackage[version=4]{mhchem}

\newcommand{\new}[1]{\textcolor{black}{#1}}

\begin{document}

\title{Microfluidic-based Bacterial Molecular Computing on a Chip}

\author{Daniel~P.~Martins,
        Michael Taynnan Barros,
        Benjamin O'Sullivan,
        Ian Seymour,
        Alan O'Riordan,
        Lee Coffey,
        Joseph Sweeney,
        and~Sasitharan~Balasubramaniam,%
\thanks{Daniel P. Martins and Sasitharan Balasubramaniam are with the Walton Institute for Information and Communication Systems Science, Waterford Institute of Technology (WIT), Waterford, Ireland, X91 P20H. E-mail: \{daniel.martins, sasi.bala\}@waltoninstitute.ie.}
\thanks{Michael Taynnan Barros is with the University of Essex, Colchester, United Kingdom, CO4 3SQ, and with the Computational Biophysics and Imaging Group, BioMediTech, Faculty of Medicine and Health Technology, Tampere University, FI-33014 Tampere, Finland. Email: m.barros@essex.ac.uk}
\thanks{Benjamin O'Sullivan, Ian Seymour and Alan O'Riordan are with Tyndall National Institute, Cork, Ireland, T12 R5CP. E-mail:\{benjamin.osullivan, ian.seymour, alan.oriordan\}@tyndall.ie}
\thanks{Lee Coffey is with the Pharmaceutical \& Molecular Biotechnology Research Centre, Waterford, Ireland, X91 K0EK. E-mail: lcoffey@wit.ie}
\thanks{Joseph Sweeney is with the LIFE farm4more, University College Dublin, Dublin, Ireland. E-mail: joseph.sweeney@ucd.ie}}


\maketitle

\begin{abstract}
Biocomputing systems based on engineered bacteria can lead to novel tools for environmental monitoring and detection of metabolic diseases. In this paper, we propose a Bacterial Molecular Computing on a Chip (BMCoC) using microfluidic and electrochemical sensing technologies. The computing can be flexibly integrated into the chip, but we focus on engineered bacterial AND Boolean logic gate and ON-OFF switch sensors that produces secondary signals to change the pH and dissolved oxygen concentrations. 
We present a prototype with experimental results that shows the electrochemical sensors can detect small pH and dissolved oxygen concentration changes created by the engineered bacterial populations' molecular signals. Additionally, we present a theoretical model analysis of the BMCoC computation reliability when subjected to unwanted effects, i.e., molecular signal delays and noise, and electrochemical sensors threshold settings that are based on either standard or blind detectors.  
Our numerical analysis found that the variations in the production delay and the molecular output signal concentration can impact on the computation reliability for the AND logic gate and ON-OFF switch. The molecular communications of synthetic engineered cells for logic gates integrated with sensing systems can lead to a new breed of biochips that can be used for numerous diagnostic applications. 
\end{abstract}


\IEEEpeerreviewmaketitle

\section{Introduction}\label{sec:introduction}

\IEEEPARstart{B}{iocomputing} is an emerging research field that envisions the use of biological componets to create computing tasks in the very same way that silicon technology is used today in conventional computing devices \cite{Moreno2019}. These systems can be based on prokaryotic cells, such as bacteria, and can be used for example applications such as detection of metal ions, as well as controlled communication functionalities through emission of signalling molecules \cite{Hsu2016,Pasotti2018,Higashikuni2017}. Biocomputing systems can be integrated into molecular computing chips, where they can be used as biosensors to diagnose and analyse biological samples and specimen. 
The computing function can be achieved through population signaling and communication, and one example is through the engineering of bacterial communications such as (\emph{quorum sensing} signalling), which is a form of communication used by microbes to coordinate functions in both small or large populations \cite{Mehta2009,Tamsir2011}. 

Bacteria signalling processes have been used in the design of engineered nanoscale biological systems for performing logic computations \cite{Din2020,Shin2020,Green2017,Martins2018,Martins2019}. Such engineered signalling processes can tune the computing operation accuracy and this builds on a new communications theory paradigm known as {\bf Molecular Communications} \cite{Martins2018,Martins2019,Akan2017,Abbasi2017,Akyildiz2019,Farsad2016,Bi2021}. The characterization and design of artificial communication systems built from biological components found in nature is the main goal for Molecular Communications systems. 
However, the development of an operational molecular computing system for diagnostics will require other accompanying technologies, such as translation from chemical into electrical signals. An example of this technology, considered in this paper, is electrochemical-based sensing. However, interfacing biocomputing cells to electrochemical sensing systems can lead to challenges that includes obtaining precise readings of the noisy biologically computed molecular signals \cite{barros2021engineering,Pierobon2011,Plesa2018}. To that end, here we propose a bacteria-based biocomputing system and design a method for the accurate detection of molecular signals emitted by it. Based on that, we evaluate the end-to-end reliability of the proposed biocomputing system within a biochip.

Biochips based on microfluidics (a.k.a, lab-on-chip) have been extensively reported in the literature, specially for the drug discovery and \emph{in situ} diagnostics. These devices are portable and easy-to-use platforms for the analysis of biomolecules and are driving the innovation in the fields of life sciences, and biochemistry \cite{Khalid2017,Valera2018,Lombardi2011,Dudala2019,Zhao2010,Yagi2007,Din2020}. 
Inspired by these works, we introduce the concept of \emph{Bacterial Molecular Computing on a Chip (BMCoC)}, which integrates electrochemical sensors and bacteria-based molecular computing systems by analyzing and estimating the molecular communication through signal detection theory.

\begin{figure*}
\centering
\subfloat[\label{fig:bmcoc_a}]{
\includegraphics[width=0.6\textwidth]{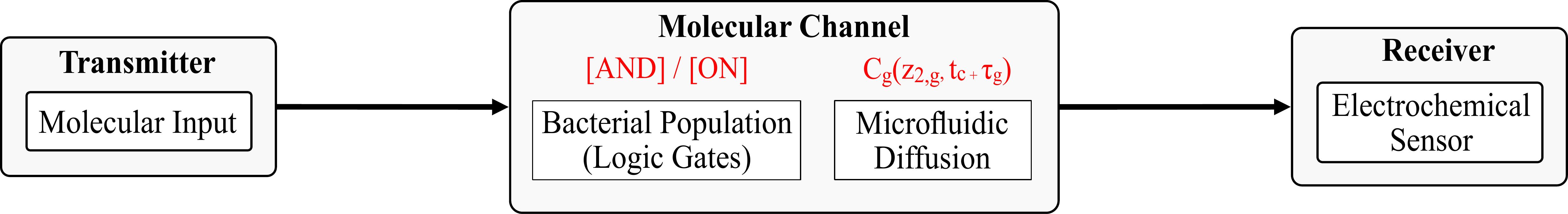}}
\vfill
\subfloat[\label{fig:bmcoc_b}]{
\includegraphics[width=\textwidth]{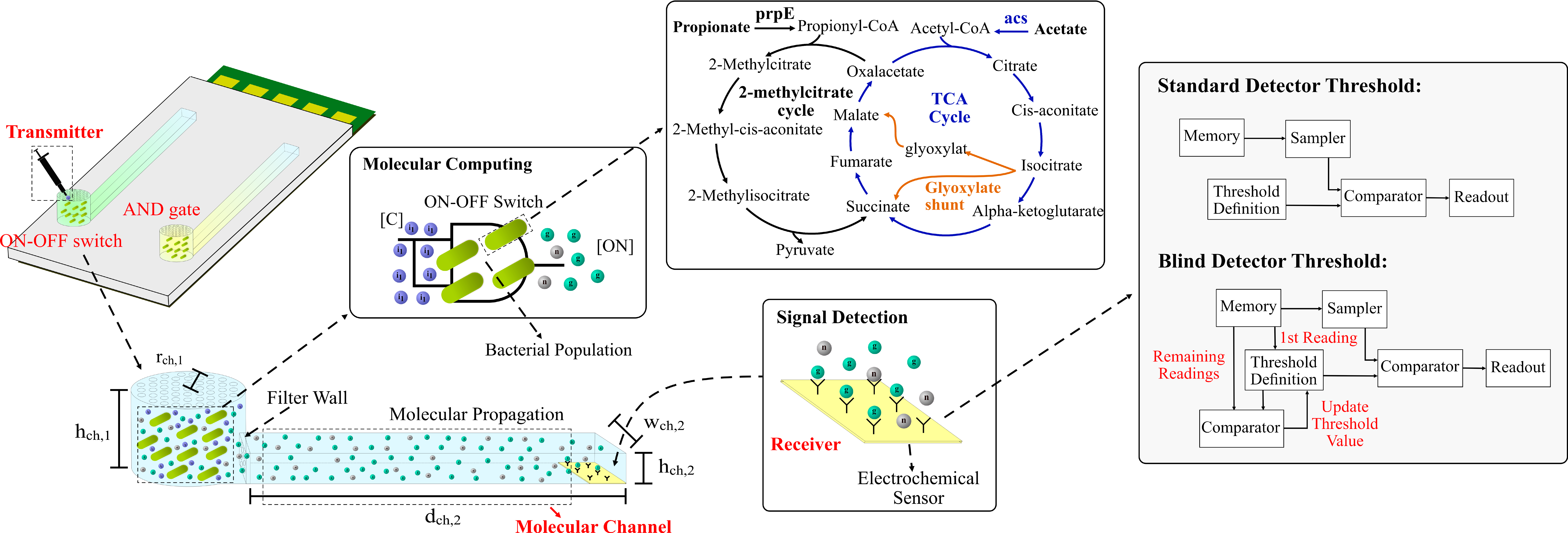}}
\caption{\new{Illustration of the integration of a bacteria-based molecular communications system with an electrochemical sensing device proposed in this paper. (a) Representation of the bacteria-based molecular communications system that supports the operation of the engineered cell computing. (b) Each bacterial population process the molecular input signal using a specific genetic engineering (represented as logic gates) and is encased in a microfluidic chamber connected to the electrochemical sensor through a microfluidic channel. The sensed molecular signal is then detected using one of the proposed estimation techniques.}}
  \label{fig:model}
\end{figure*}

Our proposed solution differs from the current lab-on-chip devices due to the molecular signal detection and computing using an AND Boolean logic gate and an ON-OFF switch sensors that is constructed from engineered bacterial populations \cite{Farsad2012,Grebenstein2019}. The molecular signal output from the biocomputing operation changes the pH or the dissolved oxygen concentration of the fluid media inside a chamber, allowing it to be detected and quantified by the electrochemical sensors. This measurement is the basis for our chip's molecular computing reliability analysis (see Figure \ref{fig:model}). We also consider other unwanted effects such as molecular input delays and signal production fluctuations that can affect the computing performance. We use a similar electrochemical sensing system design as in \cite{Din2020} to measure the pH and the dissolved oxygen concentration change due to the emission of molecular signals from the engineered bacterial population. Moreover, our suggested application improves on the design proposed by \cite{Hiyama2008} as it considers lesser moving parts and relies on free-diffusion to transport the molecular information. For our proposed system, the engineered bacterial populations receive chemicals to produce and emit molecular signals through a microfluidic channel that gets propagated towards an electrochemical sensors to detect the output from the computing operation. Our main contributions are as follows:
\begin{itemize}
\item {\bf  Analysis of the BMCoC components through wet-lab experiments:} The engineering of the bacterial populations and the electrochemical sensors are introduced and wet lab tests are performed to describe the performance of these main components of the BMCoC design.
\item {\bf  A communication system model for the analysis of the BMCoC performance:} We propose the use of multiple engineered bacterial populations to compute different molecular input signals and theoretical analysis of the communications processes and molecular output signal detection that might affect the performance of the BMCoC.  
\item {\bf  Analysing the reliability of molecular environmental signals computation:} We analyse two factors, delay and molecular input signal concentration, that can affect the reliable processing of molecular generated signals and, consequently, the computed signals of the logic gates to be analyzed by the chip.  
\end{itemize}

In the next section we introduce and provide an overview of BMCoC. In Section \ref{sec:physical_design}, we describe the physical design of the BMCoC. Then, in Section \ref{sec:mc_model} we present the molecular communications model for the BMCoC, which supports the signal detection estimation and computing reliability introduced in Section \ref{sec:estimation}. Next, in Section \ref{sec:results} we present our experimental results for  molecular computing by the bacterial population, and the results obtained for the analysis of the BMCoC reliability logic computation for varying delays and molecular input signal concentrations. Lastly, in Section \ref{sec:conclusions} we present our conclusions.

\section{Overview of the BMCoC}\label{sec:overview}

BMCoC is a device that transduces the molecular signals computed by engineered bacteria into electrical current or potential and can be connected to a wireless interface, enabling its remote monitoring. 
\new{The BMCoC contain a number of microfluidic tubes that will store and interconnect the engineered bacteria with the electrochemical sensors, which are placed on a printed circuit board. Figure \ref{fig:model} illustrates the molecular communications system proposed to support the operation of the BMCoC within microfluidic tubes and can be divided into two chambers: one that stores the engineered bacteria and the second tube where the molecules emitted by the bacterial population are diffused (see Section \ref{sec:physical_design} for more details). For the proposed device, the molecular input signals are computed by each bacterial population resulting in a secondary signal that is diffused towards the electrochemical sensor. This bacteria-based molecular communications system is represented in Figure \ref{fig:model} (a).} 

To process the molecular input signals placed on the BMCoC surface, the bacteria are engineered to act as logic gates 
(AND gate and ON-OFF switch), and the molecular output signal produced by these bacterial populations modifies the pH of the fluid media or the dissolved oxygen concentration in the microfluidic tube. \new{Figure \ref{fig:model} (b) illustrates how the BMCoC receives, processes and detects the molecular signals. The whole process starts with the insertion of the molecular input signals 
that will diffuse towards the engineered bacteria and 
produces output secondary molecular signal that will be diffused through the microfluidic tube and detected by the electrochemical sensor. The detection process is based on an estimation technique (see Figure \ref{fig:model} (b)). When all the system's characteristics are known \emph{a priori}, the electrochemical sensor circuit board implements a standard detector threshold. An alternative technique is based on the blind detector threshold that can be implemented to increase the reliability of the system. Please note that, while in the following section we further detail the physical design of the BMCoC, the detailed experimental methodology is described in the Appendix.}

\section{Physical Design of the BMCoC}\label{sec:physical_design}

\new{In this section, we introduce the physical design of the BMCoC's main components: a microfluidic structure that encase the bacterial populations and serves as the waveguide for the molecules produced by them, and the custom-made electrochemical sensors chip that will detect the pH and dissolved oxygen variation around the electrodes.}
\vspace{-.3cm}

\new{\subsection{Microfluidic Tube Structure}}
The design of the microfluidic tube follows a similar approach to \cite{Dudala2019,Dawson2014}, where polydimethylsiloxane (PMDS) is used. As mentioned in Section \ref{sec:overview}, the microfluidic tube is divided into two chambers. One chamber is shorter than the other and stores the bacterial population with a volume of $V_{ch,1}=\pi r_{ch,1}^2h_{ch,1}$, where $r_{ch,1}$ and $h_{ch,1}$ are the radius and the height of the microfluidic chamber, respectively. The longer microfluidic chamber is designed to allow the free diffusion of the molecular output signal emitted by the bacterial population, and its volume is given by $V_{ch,2}=d_{ch,2}h_{ch,2}w_{ch,2}$, where $h_{ch,2}$, $d_{ch,2}$ and $w_{ch,2}$ are the height, length and width of the chamber. The dimensions of each microfluidic tube are defined with respect to the desired applications. The two chambers that make up the tube are interconnected by an encapsulating porous membrane that allows the molecules to flow through the tube and not the bacterial cells \cite{Shim2011} \cite{Martins2019}.

\new{\subsection{Electrochemical Sensor}}
The electrochemical sensors chip situated at the bottom of the tube consists of two combs of gold working interdigitated electrodes, platinum pseudo reference and gold counter electrodes, and they are interfaced to an external electronics via a microSD port to facilitate electrical connection  \cite{Dawson2012,Barry2013,Dawson2014}. The sensors are fabricated on a 4-inch silicon wafers bearing a thermally grown $300$ nm silicon dioxide layer. Blanket metal evaporations of Titanium ($10$ nm) and Gold ($100$ nm) using a Temescal FC-2000 E-beam evaporator and lift-off technique yields interdigitated microband ($55\,\mu$m x $1\,\mu$m x $60$ nm) structures with gaps between the combs of  $1$, $2$ and $10\,\mu$m. A second metal evaporation and lift-off process yield the interconnection tracks, contact pads and the gold counter electrode ($90\,\mu$m x $7$ mm). Finally, a third metal evaporation was performed to create the platinum pseudo reference electrode that is used to detect the current magnitudes in this system. To monitor the operation of this pseudo reference electrode during this development phase, we used another reference electrode that was not built into the system allowing us to check for any reading inconsistencies. To prevent unwanted interactions along the connection tracks, silicon nitride, which acts as an insulating layer was deposited by plasma-enhanced chemical vapour deposition. Photolithography and dry etching were utilised to selectively open windows ($45\,\mu$m x $100\,\mu$m) in the insulating SiN layer over the microband electrodes for electrolyte access. Openings were also created over the counter and pseudo-reference electrodes and the contact pads. Each device contains six interdigitated electrode sensors which are separated by $0.94$ mm. Once the sensor fabrication is completed, a wafer was diced into $28$ separate chip devices.  

A custom-made holder cell was fabricated to allow measurement in small electrolyte volumes ($\approx50\,\mu$L to $5$ mLs). The cell was constructed from an aluminium base and a Teflon lid. Spring-loaded probes (Coda Systems Ltd. PM4J Plain Radius Microprobes) were inserted into the lid in position above the peripheral contact pads, to permit electrical connection to external potentiostats. The cell was assembled with a Viton O-ring embedded in the lid to form a seal around the on-chip electrodes. Viton O-rings were chosen for their chemical resistance. The inner diameter of the O-ring was $7$ mm with a cross-section of $1.6$ mm to allow an opening large enough to expose all six sensors, counter and reference electrodes on the device to the electrolyte.

\vspace{.3cm}
\section{MC Model of the BMCoC}\label{sec:mc_model}



\new{In this paper, we model the different components of the BMCoC using molecular communications theory, and the represented communication model is illustrated in Figure \ref{fig:model} (a). For this model, we are especially interested in describing the insertion, processing and detection of molecules. A typical molecular communications system have a molecular transmitter and receiver interconnected by a communications channel \cite{Akan2017, Akyildiz2019}. Therefore, we model the process of inserting a molecular signal into the BMCoC as the transmitter, the engineered bacteria and the microfluidic propagation chamber as the channel and the electrochemical sensor as the receiver. Figure \ref{fig:model} (b) shows the two engineered bacterial gates that produces molecules that propagate through the microfluidic tube channel.} 


\subsection{Transmitter}

The transmitters considered in this communications system is the 
engineered bacterial logic gates that will get triggered once molecular input signals are inserted into the cell population. The molecular input signals are here assumed to be a train of pulses with amplitude $m_i$ and are defined as follows \cite{Llaster2013}
\begin{equation}
[i]=\dfrac{m_is_i}{\sqrt{4\pi D(t_c+\tau_{in})}}e^{\dfrac{-z_{1,g}^2}{4D(t_c+\tau_{in})}}
\end{equation}
where $s_i$ is the representation of the pulses generated by the insertion of the molecules on the BMCoC, $m_i$ is the concentration (pulse amplitude) of the molecular input signals, \new{$z_{1,g}$ is the Euclidean distance from the insertion point and the centre of the bacterial population}; $\tau_{in}$ is the propagation delay between insertion point and the bacterial population, $i=\{A,B,C\}$ is the representation of the three molecular input signals considered for the operation of the AND gate and ON-OFF switch, $g=\{AND, ON\}$ is the representation of the different molecular logic operation, and \new{$t_c$ is the duration of the molecular signal's insertion into the bacterial population's chamber.}

\subsection{Channel}\label{sec:theo_model}
Bacteria can process molecular signals as switches, amplifiers, or logic gates \cite{Akan2017,Martins2019}. In this case, the mathematical formulations model the chemical reactions that consume or transform the molecular concentrations used by these synthetic processes. 
While the ON-OFF switch operates using a single molecular concentration to produce a secondary signal that will affect the dissolved oxygen concentration around the electrochemical sensors, the AND gate requires two molecular input signals to generate a molecular concentration output that will modify the pH of the fluid. The following are the chemical reactions for each of the logic gates and is represented as follows 
\begin{equation}\label{eq:ANDgate}
\begin{split}
\dfrac{d[AND]}{dt}=&\dfrac{[A]^{n}}{K_{A}^{n}+[A]^{n}}\cdot\dfrac{[B]^{n}}{K_{B}^{n}+[B]^{n}}\\[2mm]
&-\gamma[AND]+N_{AND}(t)
\end{split}
\end{equation}
\begin{equation}\label{eq:ON_OFF}
\begin{split}
\dfrac{d[ON]}{dt}&=\dfrac{([C]^n)^2}{(K_{C}^n)^2+2K_{C}^{n}[C]^{n}+([C]^{n})^2}\\[2mm]
&-\gamma[ON]+N_{ON}(t),
\end{split}
\end{equation}
where $K_{A}$, $K_{B}$ and $K_{C}$ are the association constants for the signals $[A]$, $[B]$ and $[C]$; $\gamma$ is the consumption rate of the molecular output signal; $n$ is the Hill coefficients for these molecular signals, and $N_{AND}(t)$ and $N_{ON}(t)$ is the fluctuation of the molecular signal production which is modelled as an AWGN noise for these molecular output signals (see \cite{Martins2019} for a further explanation about this assumption). 

The molecules produced by each biological entity considered in this paper will travel through the microfluidic tube independently. Therefore, the communications channel $C_{g}(z_{2,g},t_c+\tau_g)$ can be  characterised by using the solutions for fluid media which is the Fick's diffusion equation \cite{Llaster2013}, and represented as follows \cite{Llaster2013}
\begin{equation}\label{eq:hz1}
C_{g}(z_{2,g},t_c+\tau_g)=\dfrac{1}{\sqrt{4\pi D(t_c+\tau_g)}}e^{\dfrac{-z_{2,g}^2}{4D(t_c+\tau_g)}},
\end{equation}
where $D$ is the diffusion coefficient for the propagation of the molecular signals in the fluid channels; $\tau_g$ is the propagation delay between the bacterial population and the electrochemical sensor; \new{$z_{2,g}$ is the Euclidean distance between the engineered bacterial gates centre and the electrochemical sensor}. This distance assumption ensures that each bacterium will equally contribute to the molecular computation. \new{This model also assume that the molecules dimensions are much smaller than the microfluidic tubes dimensions, allowing us to model the channel using a solution for Fick's diffusion equation.}

\subsection{Receiver}
The molecular signal produced by these engineered bacterial populations are then propagated through the channel $C_{g}(z_{2,g},t_c+\tau_g)$, resulting in the molecular signal $y_g(z_{2,g},t_c+\tau_g)$ that reach the region where the electrochemical sensors are able to detect the changes in pH or the dissolved oxygen concentration. The molecular signal $y_g(z_{2,g},t_c+\tau_g)$ can be evaluated as follows
\begin{equation}\label{eq:y}
y_g(z_{2,g},t_c+\tau_g)=[g]*C_{g}(z_{2,g},t_c+\tau_g),
\end{equation}
where $*$ denotes the convolution operation, and $[g]$ is the molecular signal concentration produced by the gates $g=\{AND,ON\}$, which is evaluated using (\ref{eq:ANDgate}) and (\ref{eq:ON_OFF}). The molecular output signal that is detected by the electrochemical sensor is given by
\begin{equation}
y_f(z_{2,g},t_c+\tau_g)=y_g(z_{2,g},t_c+\tau_g)+n_g(z_{2,g},t_c+\tau_g)
\end{equation}
where $n_g(t)$ is the electrolyte noise that affect the detection process by the electrochemical sensor and is defined as \cite{Deen2006}
\begin{equation}
n_g(z_{2,g},t_c+\tau_g)=4kTR_b(z_{2,g},t_c+\tau_g),
\end{equation}
where $R_b(z_{2,g},t_c+\tau_g)=\dfrac{1}{\Gamma(z_{2,g},t_c+\tau_g)}\sqrt{\dfrac{\pi}{a_e}}$ is the resistive process due to the passage of the molecular signal on the electrochemical sensors, $\Gamma(z_{2,g},t_c+\tau_g)=\Gamma_sy_g(z_{2,g},t_c+\tau_g)\times10^3$ is the conductivity of the molecular signal \cite{Atkins2006}, $k$ is the Boltzmann constant, $T$ is the absolute temperature, and $a_e$ is the passage area of the electrochemical sensors. 


\section{Estimation of Molecular Signal Detection}\label{sec:estimation}

\new{Existing channel properties affect the signals that reach the electrochemical sensors, including delays and noise, and impact on the quality and capacity of a molecular communications system \cite{Pierobon2011,Akkaya2014,Martins2019}. Therefore, we consider two signal reception estimation techniques (standard and blind detector thresholds) to ensure that the electrochemical sensors is able to detect lower levels of molecular signals and their respective concentration values. 
Our goal is to verify whether these estimation techniques can improve the reliability of the logic computation.}

\subsubsection{Standard Detector Threshold}

\new{We first consider the case where the electrochemical sensors know \emph{a priori} the characteristics of the microfluidic tube, the engineered bacterial population and the system inputs. In this case, a standard detector threshold, $r_{std,g}$, is defined based on the molecular output signal concentration that the engineered bacteria will diffuse through the microfluidic tube. This is defined
as follows
\begin{equation}\label{eq:r_std}
\hat{r}_{std,g}=\dfrac{\max(y_f(z_{2,g},t_c+\tau_g))}{2},\,\,\text{for}\quad0\leq t_c\leq t_p,
\end{equation}
where $t_p$ is the period of each pulse.} 

\subsubsection{Blind Detector Threshold}

\begin{figure*}
\centering
\subfloat[\label{fig:wetlab_a}]{
\includegraphics[width=0.3\textwidth]{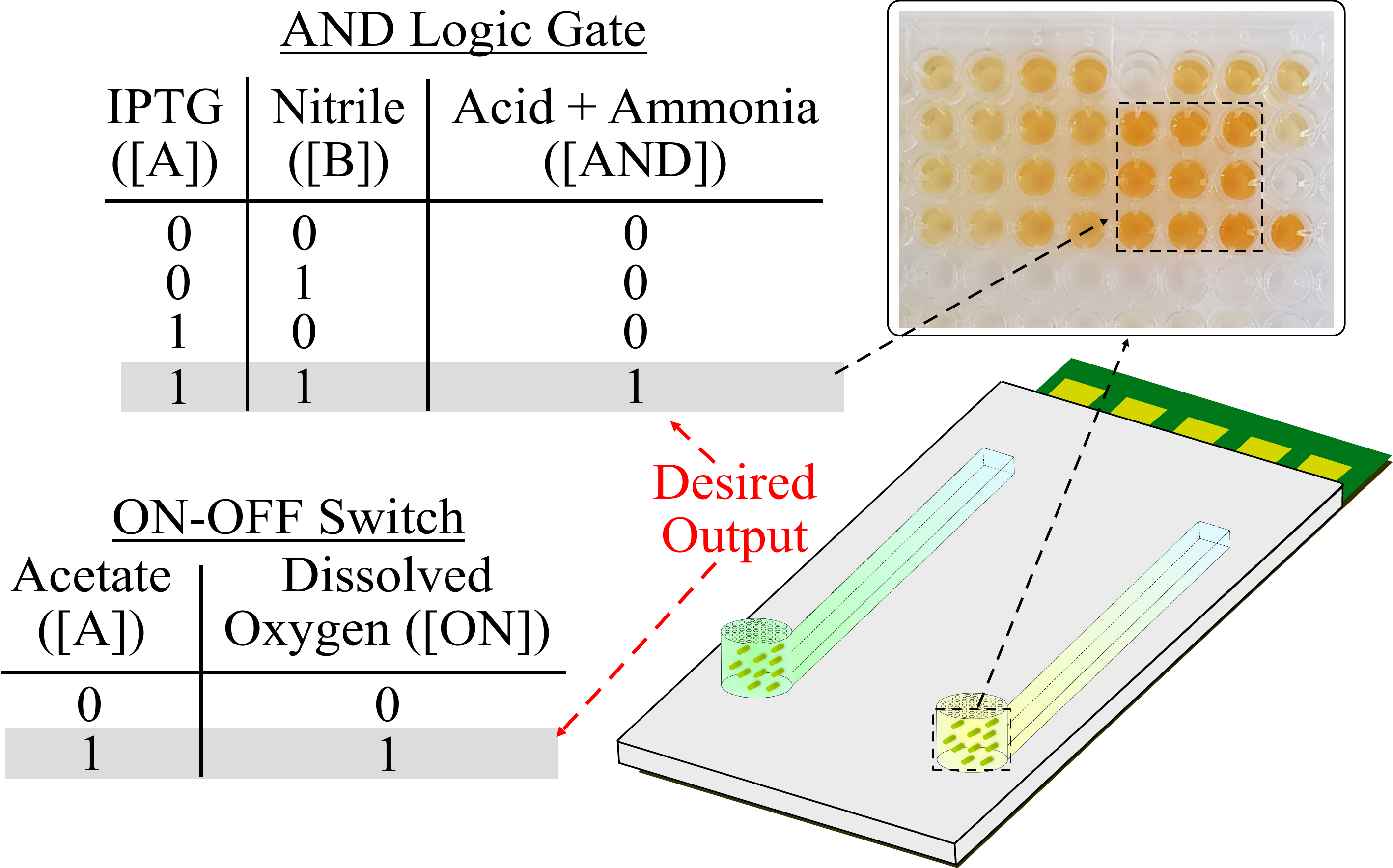}}
\hfill
\subfloat[\label{fig:wetlab_b}]{
\includegraphics[width=0.3\textwidth]{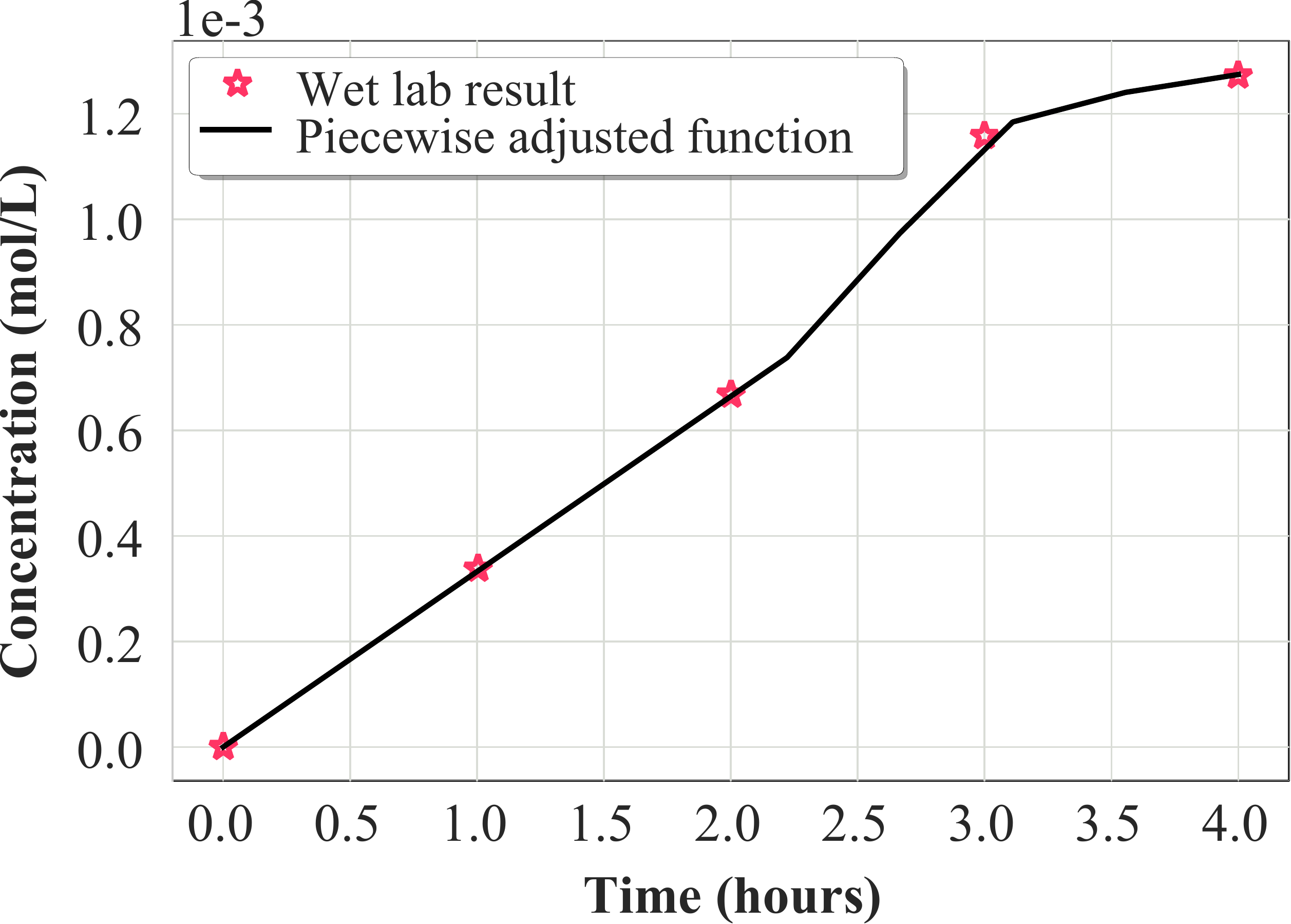}}
\hfill
\subfloat[\label{fig:wetlab_c}]{
\includegraphics[width=0.35\textwidth]{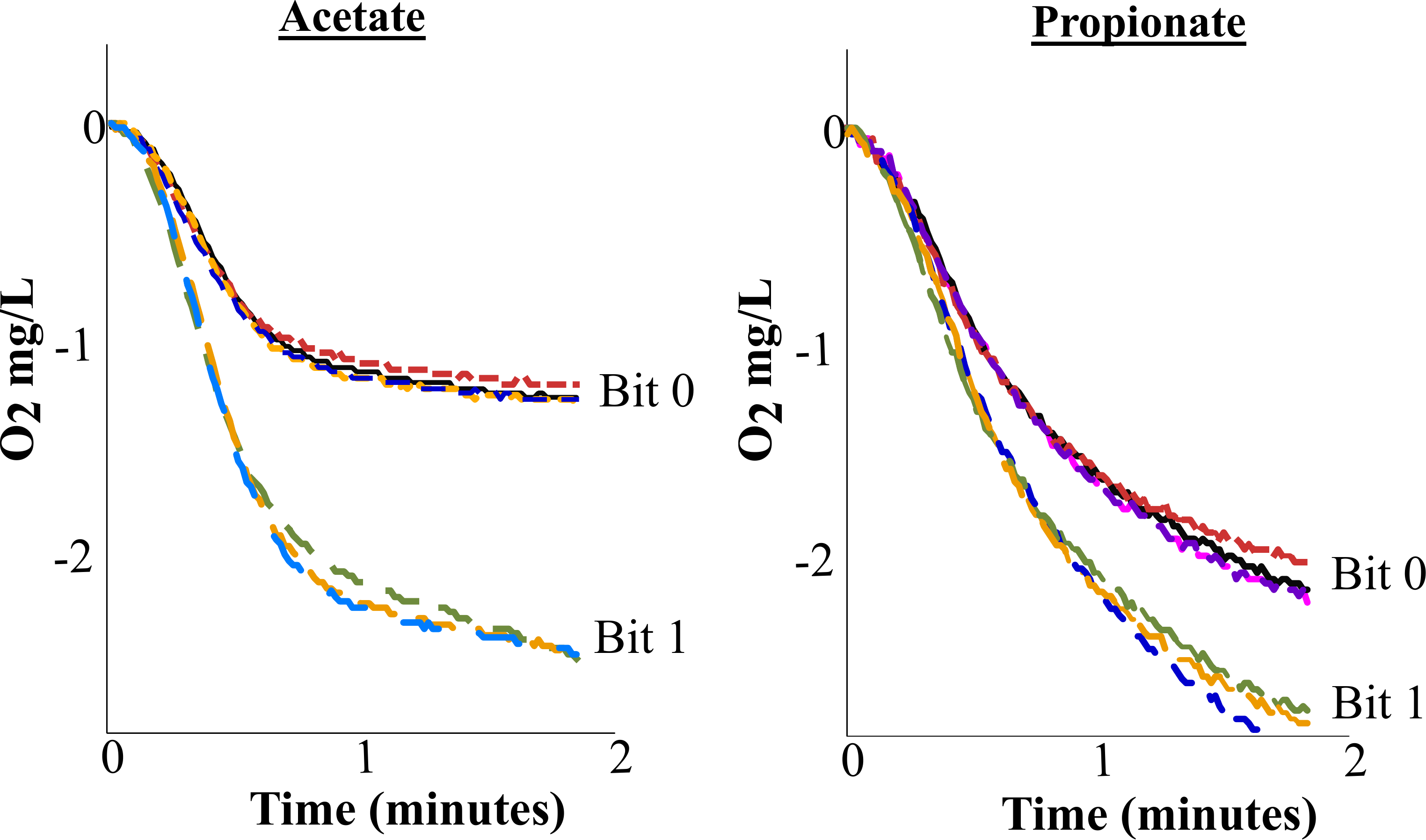}}
\caption{\new{Illustration of the wetlab experiments performed to investigate the performance of the bacterial populations. (a) A 96-well plate assay was performed for the synchronized production of the molecular signal by the AND gate. (b) Linear fitting of the molecular concentration produced by the bacterial population is shown for the desired output of the AND gate. (c) Plot results for the acetate and propionate exclusion ON-OFF switches operation. For the acetate case, the ON-OFF switch shows a $\text{O}_2$ consumption response for $0.2$ mmol/l and $0.4$ mmol/l molecular input (bit ``0'' and bit ``1'', respectively), while the propionate ON-OFF switch shows its $\text{O}_2$ consumption responses for $0.16$ mmol/l and $0.21$ mmol/l (bit ``0'' and bit ``1'', respectively).}}
\label{fig:wetlab}
\end{figure*}

For the blind detector case, the electrochemical sensors do not know \emph{a priori} the molecular output signal concentration that the bacterial population can generate \cite{Dabiri2017}. The BMCoC device will continuously read the pH or the dissolved oxygen concentration change and store information for \emph{a posteriori} detection. The proposed estimator, inspired by \cite{Dabiri2017}, consists of two steps. First, the electrochemical sensor will measure the maximum molecular output signal concentration, $y_l$, produced for a short period, $L_p$. Then the initial detection threshold value is defined as follows 
\begin{equation}\label{eq:tholdblind_init}
\hat{r}_{1,g}=\dfrac{\max(y_f(z_{2,g},t_c+\tau_g))}{L_p},\quad \text{for}\quad0\leq t\leq t_p,
\end{equation}
where $L_p$ is the short period that the sensor would measure the molecular concentration. Second, the electrochemical sensor will continue to measure the molecular output signal concentration to improve the detection threshold. In this stage, the electrochemical sensor will compare the defined threshold for the first pulse with the maximum concentration measured in the next pulse. If the result lies below $0.5$ (to be closer to the standard detector threshold value), the electrochemical sensor adjusts the blind detector threshold; otherwise, it maintains the previous threshold value. We describe this process as follows
\begin{equation}
\hat{r}_{r,g}=\dfrac{\hat{r}_{1,g}}{\max(y_f(z_{2,g},t_c+\tau_g))},\quad \text{for}\quad0\leq t\leq t_p.
\end{equation}
By using the value of the parameter $\hat{r}_r$, the electrochemical sensor would define whether the threshold should be adjusted or not. Therefore,
\begin{equation}\label{eq:tholdblind_end}
\begin{cases}
\hat{r}_{r,g}&<0.5,\,\hat{r}_1\,\text{will be increased}\\
\hat{r}_{r,g}&\geq0.5,\,\hat{r}_1\,\text{will be maintained}.
\end{cases}
\end{equation}
This adjustment process will take place in every reading after the first evaluation of the threshold. 

\subsection{Reliability Analysis of Logic Computation}

\new{Uncertainties, noises and delays, affecting the signal produced by the bacterial populations might result in the incorrect detection of the emitted pulses by the electrochemical sensors. In typical communications systems, the probability of error is often used to evaluate the impact caused by these uncertainties and considered as a performance metric  \cite{Pierobon2011,Martins2019}.  Here we quantify the accuracy of the BMCoC molecular computations by investigating the probability of obtaining correct values from the detection. To measure the reliable logic computation probability ($RLC$) of the BMCoC, we first sampled a received signal to evaluate the number of ``0's'' and ``1's'' that are correctly detected. The digitalisation process is defined as follows}
\begin{equation}\label{eq:digital_detect}
\begin{cases}
y_f(z_{2,g},t_c+\tau_g)\geq\hat{r}_{std,g},\,\, y_f[j]=1\\
y_f(z_{2,g},t_c+\tau_g)<\hat{r}_{std,g},\,\, y_f[j]=0\\
y_f(z_{2,g},t_c+\tau_g)\geq\hat{r}_{r,g},\,\, y_f[j]=1\\
y_f(z_{2,g},t_c+\tau_g)<\hat{r}_{r,g},\,\, y_f[j]=0,
\end{cases}
\end{equation}  
where each pulse is composed of 50 samples $j$ and $j_{tot}=500$ is the total number of samples. Based on the digital representation of the molecular output signal, we define the reliable logic computation $RLC$ as the measurement of the number of correct detections performed by the electrochemical sensor. In other words, we can compute how many ``0's'' and ``1's'' are correctly defined (true negatives - $TN$, and true positives - $TP$) with respect to the total detected molecular output signal. Therefore, we describe the probability of reliable logic computation as 
\begin{equation}\label{eq:rlc}
RLC=\dfrac{TP+TN}{j_{tot}}\times100.
\end{equation}

\begin{figure}[t!]
  \centering
  \includegraphics[width=\columnwidth]{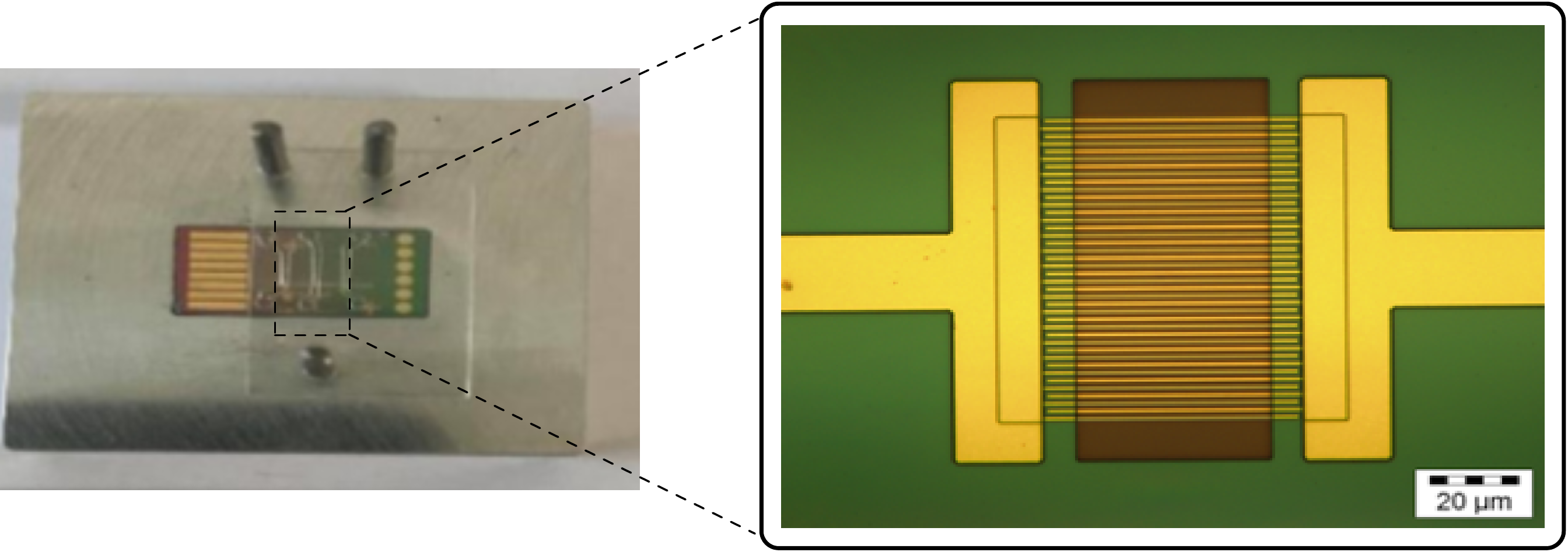}
  \caption{\new{Image taken of the microfluidic channel and electrodes considered in this paper. We magnified ($50x$) a section of the microfluidic channel to visualise the gold IDE array used for both pH measurements and oxygen quantification.}}
  \label{fig:microfluidic}
 \end{figure}


\section{Results}\label{sec:results}

\new{In this section, we present the results from our performance analysis of the BMCoC. First, we introduce the results of the molecular signals production by the engineered bacterial populations, AND gate and ON-OFF switch. Second, we present the results of the electrochemical sensing process, including the generated electrical current and potential caused by the pH and dissolved oxygen changes. Finally, we present the results of the reliable logic computation probability when using the standard and blind detector thresholds and the molecular communications systems is affected by noises and delays, from the molecular production and propagation.} 

\begin{figure*}[t!]
  \centering
  \subfloat[\label{fig:ph_results}]{
  \includegraphics[width=0.4\textwidth]{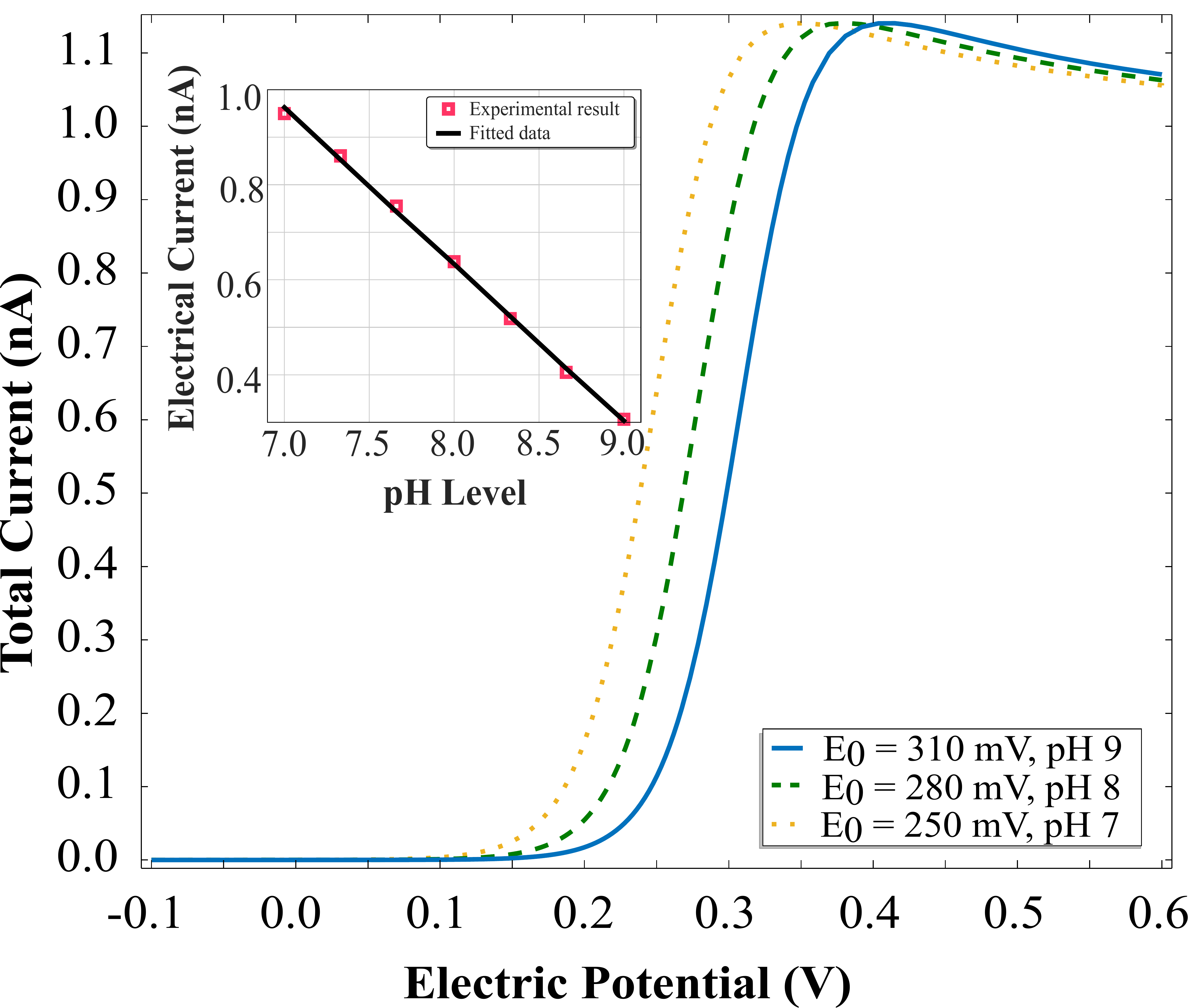}}
  \hfill
  \subfloat[\label{fig:o2_results}]{
  \includegraphics[width=0.48\textwidth]{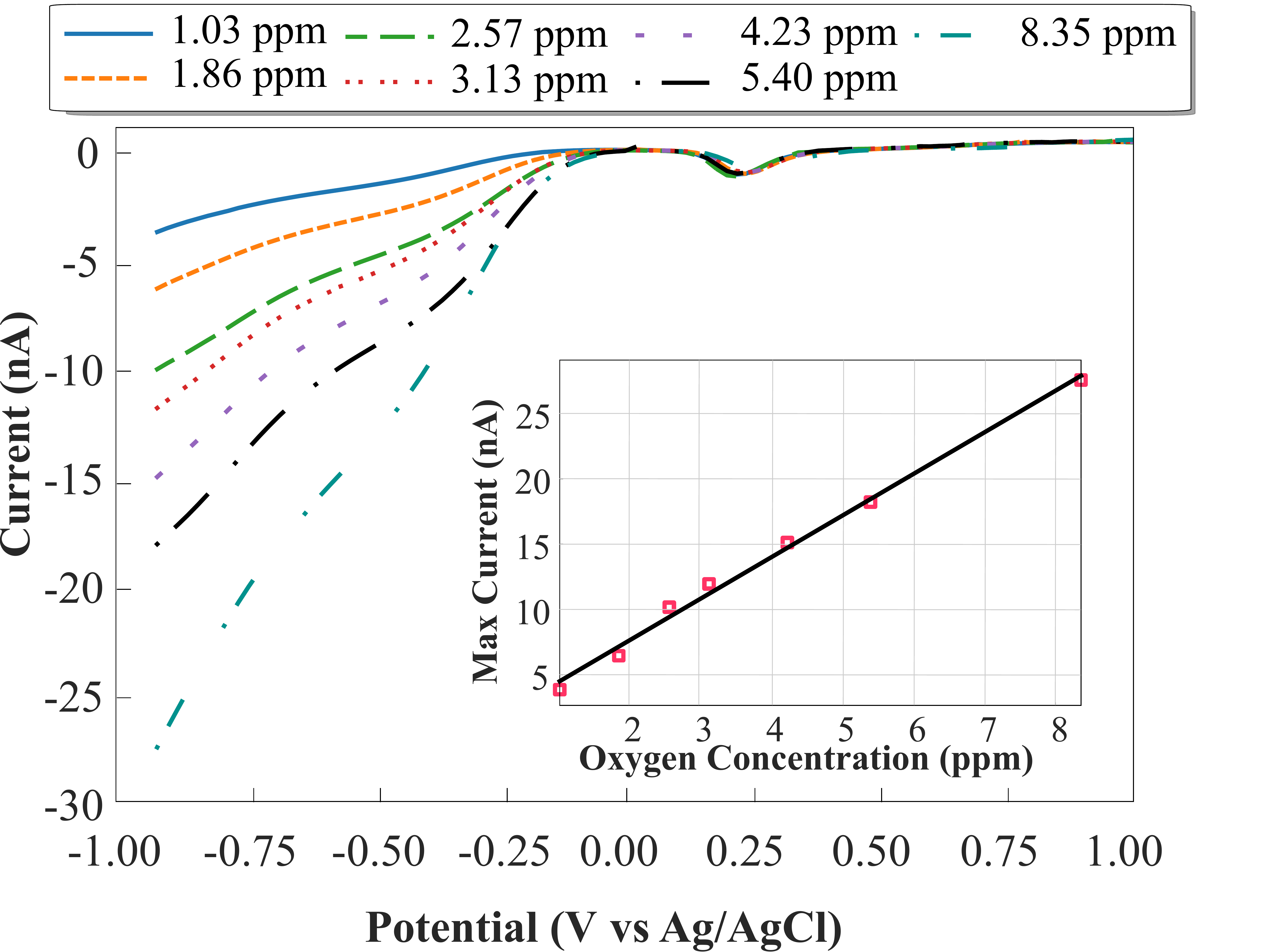}}
  \caption{\new{Investigation of the operation range for the electrochemical sensors. (a) LARGE: Cyclic voltammograms over the pH range of $7$ to $9$. As the pH decreases (as more molecules are generated), a lower redox potential $E_0$ of ferrocene is needed. SMALL: The electrical current values for different pH levels when the electric potential is fixed at $0.28\,\text{V}$. The decrease rate was measured as $-32.97\,\text{mV}/\text{pH}$. The sensitivity of the nanowire sensors \cite{Dawson2014} means the current value at $0.28\,\text{V}$ can be used as a probe for the pH change in a solution. (b) LARGE: Cyclic voltammograms of various concentrations (in parts per million) of oxygen at a gold microband array. The cyclic voltammograms were swept from $1.2$ V to $-0.9$ V at $50$ mV/s in water samples at pH $8$. SMALL: The calibration plot using the current values at $-0.9$ V versus the measured concentration of oxygen in water.}}
  \label{fig:electrochem_results}
\end{figure*}

\subsection{System-Wide Analysis of the BMCoC}

\subsubsection{Implementation of Bacteria Molecular Computing Mechanisms}
 
\new{In the experiment illustrated in Figure \ref{fig:wetlab},  we inserted two molecular input signals with the concentrations of $0.1\,\text{mmol/L}$ and $0.5\,\text{mmol/L}$ into a tube containing the engineered bacterial population representing the AND gate and measured the molecular output signal.  Figure \ref{fig:wetlab} (a) shows the results of the possible combinations of molecular input signals, and the different shades of yellow are the output from the AND logic operation. In this case, no colour means low molecular input signals; the light yellow represents the combination of a low and a high molecular input signals, and the darker yellow occurred when adding two high molecular input signals to the tube. As the dark yellow results from the AND gate activation, we quantified the production of the molecular output signal in this scenario for four hours, see Figure \ref{fig:wetlab} (b). Please note that the AND gate's molecular output signal increases almost linearly with time. This result was due to the chosen observation time that occurs before the engineered bacteria started operating in the steady-state regime \cite{Atkins2006}. Furthermore, this enabled us to observe how quickly the engineered bacterial population could emit the molecular output signal and help us to set up the detector threshold values applied in Section \ref{sec:theo_analysis}.}

\new{To quantify the operation of the ON-OFF switch, we engineered a bacterial population to use acetate and propionate, with $0.2$ mmol/L and $0.4$ mmol/L acetate and $0.16$ mmol/L and $0.2$ mmol/L propionate concentrations (in at least triplicate), respectively. Initial $\text{O}_2$ consumption rates ($\text{mg}.\text{O}_2.\text{min}^{-1}$) were calculated from the respective acetate and propionate $\text{O}_2$ consumption responses and presented in Figures \ref{fig:wetlab} (c). In the case of the acetate, the engineered bacteria produced mean $\text{O}_2$ consumption rates of $1.622\,\text{mg}.\text{O}_2.\text{min}^{-1}$ for $0.2$ mmol/L and  $2.967\,\text{mg}.\text{O}_2.\text{min}^{-1}$ for $0.4$ mmol/L acetate while  the propionate scenario produced $1.8$ and $2.499\,\text{mg}.\text{O}_2.\text{min}^{-1}$ for $0.16$ mmol/L and $0.21$ mmol/L propionate concentrations, respectively, see Figure \ref{fig:wetlab} (c). It is evident from both scenarios that the $\text{O}_2$ consumption rates can be applied as the trigger for the bacteria processing of acetate and propionate concentrations.}

\subsubsection{Electrochemical Sensing}\label{sec:electrochem}

We simulate the electrochemical sensor detection process using a finite element software, COMSOL Multiphysics\textregistered (version 5.3). For this particular simulation, we considered a microfluidic channel as shown in \new{Figure \ref{fig:microfluidic} (see Appendix B for more details).}

Due to the small concentration of protons (in the range of nmol/L) produced by the engineered bacterial population, the acidification of the fluid medium will be more sensitive in the pH range of 9 to 7 (from $1\,\text{nmol/L}$ to $100\,\text{nmol/L}$). For example, the standard detector threshold for the AND gate operation result in the addition of $2.27\times10^{-8}$ protons to the fluid (or pH 7.64), see Section \ref{sec:std_thold_analysis} for details. By assuming a 1:1 molecular concentration relationship, this threshold value would reduce the $\text{pH}_f$ of a fluid as follows
 \begin{equation}\label{eq:pH7.62}
 \begin{split}
 &\text{pH}_f = -\log_{10}(\text{pH}_9+\text{pH}_{7.64})\\
 &= -\log_{10}(1\times10^{-9}+2.27\times10^{-8}) = 7.62,
 \end{split}
 \end{equation}
 and
 \begin{equation}\label{eq:pH6.91}
 \begin{split}
 &\text{pH}_f = -\log_{10}(\text{pH}_7+\text{pH}_{7.64})\\
 &= -\log_{10}(100\times10^{-9}+2.27\times10^{-8}) = 6.91.
 \end{split}
 \end{equation}
 
We noted from (\ref{eq:pH7.62}) and (\ref{eq:pH6.91}) that a molecular output signal with the concentration of $2.27\times10^{-8}$mol/L can produce a greater pH level change if the fluid has a pH 9 than if it had a pH 7. These pH level changes result in the production of different electrical currents when measured at a given electric potential, as shown in \new{Figure \ref{fig:electrochem_results} (a).} A small increase in the pH level results in a significantly different electric current value when measured at a fixed potential of $0.28\,\text{V}$, facilitating the detection of small output molecular signal concentrations.

 
Based on our previous analysis, we defined an electric potential of $0.28\,\text{V}$ to simulate the detection of the molecular output signal by the electrochemical sensors. Through this experiment, we determine the electric current required to oxidate the signalling molecule (FcCOOH). \new{Figure \ref{fig:electrochem_results} (a)} show the result obtained from this analysis. When the fluid channel has a pH of 7, a higher electrical current passes through the electrochemical sensors at the fixed potential than when it has a pH of 9. Therefore, a small ion concentration change, such as $22.7\,\text{nmol/L}$, is harder to be detected for a fluid with a lower pH level. From this result, we also can propose a linear equation to fit the data and predict the electric current for other pH levels, \new{see the small plot in Figure \ref{fig:electrochem_results} (a),} as follows
 \begin{equation}\label{eq:I_c}
 I_c = -0.3219\text{pH}_c+3.1867,
 \end{equation}
where $\text{pH}_c$ is the pH level of the considered fluid media. Using (\ref{eq:I_c}), we found that for the defined threshold value ($3.2\,\text{nM}$, or pH 8.5), the electrical current produced by the electrochemical sensors was equal to $I_c=0.45\,\text{nA}$.

\begin{table}[t!]
  \centering
  \caption{Parameters considered for the theoretical analysis of the BMCoC}
  \label{tb:setup}
  \renewcommand{\arraystretch}{1.5}
  \begin{threeparttable}
  \begin{tabular}{ l  c  c c}
    \hline
    {\bf Variable} & {\bf Value} & {\bf Unit} & {\bf Reference}\\ \hline
    $K_{A}$, $K_{B}$, $K_{C}$ & $10$ & -- & \cite{Martins2019}\\
    $z_{1,g}$ & $5$ & $\mu m$ & \tnote{*}\\
    $z_{2,g}$ & $50$ & $\mu m$ & \tnote{*}\\
    $r_{ch}$ & $5$ & $\mu$m & \tnote{*}\\
    $h_{ch,1}$ & $10$ & $\mu$m & \tnote{*}\\
    $m_{A}$, $m_{C}$ & $1.2$ & mmol/L& \tnote{*}\\
    $m_{B}$ & $1.8$ & mmol/L & \tnote{*}\\
    $D$ & $1.37\times10^{-7}$ & $\text{m}^2/\text{s}$ & \cite{Martins2019}\\
    $\gamma$ & $0.01$ & -- & \cite{Martins2019}\\
    $n$ & $2$ & -- & \cite{Martins2019}\\
    $t_c$ & $720$ & seconds & \tnote{*}\\
    $t$ & $5$ & hours & \tnote{*}\\
    $t_p$ & $30$ & minutes & \tnote{*}\\
    $\tau_{in}$, $\tau_g$ & $100$ & seconds & \tnote{*}\\
    $\Gamma_s$ & $34.892$ & -- & \cite[Ch. 21]{Atkins2006}\\
    $T$ & $300.15$ & K & \tnote{*}\\
    $a_e$ & $100$ & $\mu\text{m}^2$ & \tnote{*}\\
    $k$ & $1.380649\times10^{-23}$ & J/K & \cite{SIunit2019}\\
    \hline
    \end{tabular}
    \begin{tablenotes}
    \item[*] Value extracted from the experiments by the authors.
    \end{tablenotes}
    \end{threeparttable}
\end{table}

\new{Figure \ref{fig:electrochem_results} (b) shows the measurement of the dissolved oxygen in water using a gold microband array. This figure shows the cyclicvoltammogram (CV) performed in each concentration of oxygen sweeping from $1.2$ V to – $0.8$ V vs. Ag/AgCl at $50$ mV/s for $3$ cycles.} The oxidation event at $0.7$ V corresponds to the formation of a gold oxide surface layer. The reduction event at $0.25$ V corresponds to the reduction of the formed gold oxide layer. These events are independent of oxygen, and the onset of oxygen reduction occurs at $-0.1$ V in the CV and is characterised as two waves. The first is the reduction of oxygen to hydrogen peroxide, seen at approximately $-0.5$ V. The second wave is the reduction of the hydrogen peroxide to water, seen at $-0.9$ V. These waves indicates the complete oxygen reduction, and the electric current measured is linearly dependent on the concentration of oxygen.


\begin{figure*}[t!]
 \centering
 \includegraphics[width=0.9\textwidth]{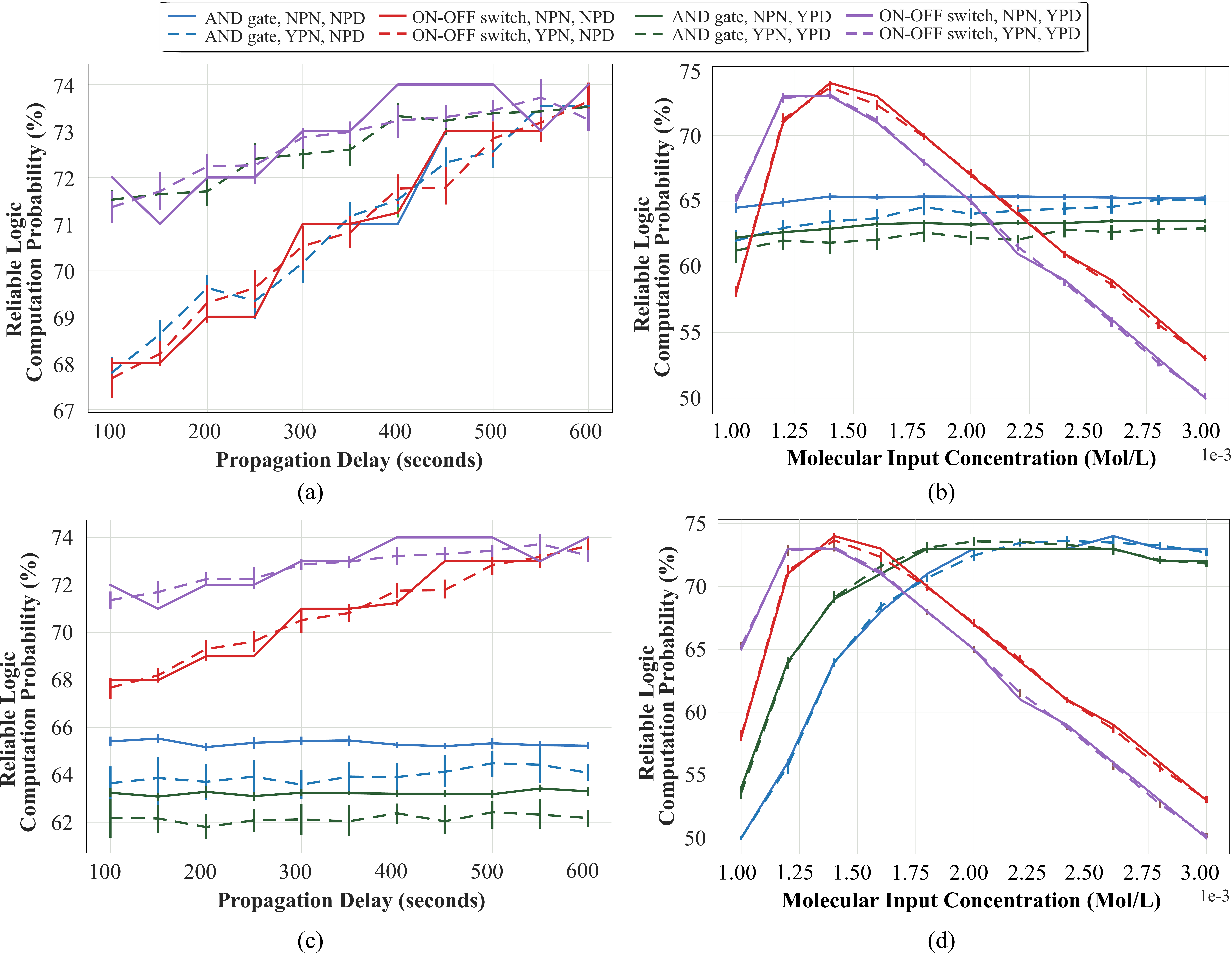}
 \caption{\new{Reliable logic computation probability for different concentrations of the molecular generated signals. (a) When the system is both affected/not affected by the production and propagation noise, as well as production delay for the AND gate and ON-OFF switch, considering a standard detector threshold. (b) Scenario similar to (a), but considering a range of molecular input concentrations. (c) When the system is both affected/not affected by the production and propagation noise, as well as production delay for the AND gate and ON-OFF switch, considering the blind detector threshold. (d) Scenario similar to (c), but considering a range of molecular input concentrations.}}
 \label{fig:theo_results}
 \end{figure*}

\subsection{Reliability of Logic Computation with Different Signal Detectors}\label{sec:theo_analysis}

Our analysis considers that the different molecular input signals inserted into each microfluidic chamber will propagate through the fluid channel (it can suffer a propagation delay $\tau_{in}$) and are detected and processed by the bacteria-based logic gate populations. Here, we apply the molecular computing models introduced in Section \ref{sec:mc_model}. Based on these models, we evaluate the reliable logic computation probability when the bacteria-based logic gate processing is instantaneous or delayed by $720$ seconds (defined as production delay). We determined the different scenarios as NPN - no production noise, NPD - no production delay, YPN - with production noise, and YPD - with production delay. Moreover, due to the stochastic nature of some processes here investigated, we run each scenario ten times and evaluate the average and standard deviation of these values. For our analysis, we use the values presented in Table \ref{tb:setup}. Additionally, we consider an AWGN noise to model the fluctuation in the production of the molecular signals.  This molecular production noise will have an average of $\mu_c=0$ and variance of $\sigma_{AND}=2\,\text{nmol/L}$ for the AND gate and $\sigma_{ON}=1\,\text{nmol/L}$ for the ON-OFF switch. The molecular output signal $y_f(h_{2,l},t_c+\tau_g)$ is produced for five hours. The bacteria-based logic gates and the electrochemical sensors have a distance of $d_{TS}=50\,\mu\text{m}$ between each other. The production and electrolyte noises, as well as the production and propagation delays, will induce errors in the molecular output signal detection. This is due to the incorrect identification of positive and negative samples in the digitalised version of the molecular output signal $y_f(h_{2,l},t_c+\tau_g)$. Therefore, we investigate the impact of the delays and noises on this bacteria-based molecular communications system by evaluating the reliability logic computation probability for different scenarios.


\subsubsection{Standard Detector Threshold Analysis}\label{sec:std_thold_analysis}

First, we investigate the electrochemical sensor standard detector threshold to study the probability of the correct detection of the molecular output signal. In this case, we considered the molecular input signals $s_A=s_B=s_C$ as pulse trains, with concentrations equal to $m_A=m_C=1.2$ mmol/L, and $m_B=1.8$ mmol/L. By applying these molecular signals in (\ref{eq:ANDgate})-(\ref{eq:r_std}), we are able to evaluate the standard detector threshold as $\hat{r}_{std,AND}=2.27\times10^{-8}$ mol/L for the AND gate and $\hat{r}_{std,ON}=1.01\times10^{-8}$ mol/L for the ON-OFF switch. We apply these standard detector thresholds in (\ref{eq:digital_detect}), evaluate the number of positive and negative samples and compare it with the molecular input signals for a varying propagation delay. Figure \ref{fig:theo_results} (a) shows the logic computation reliability for this case, and it can be noted that the AND gate is impacted more by the uncertainties of these bacteria-based communications systems than the ON-OFF switch. It can also be noted that the AND gate has a higher reliability than the ON-OFF switch for most values of propagation delay. Furthermore, as we increase the propagation delay from $100$ to $600$ seconds, the reliability logic computation probability for both AND and ON-OFF switch improves, which allow us to induce that the propagation delay counter the effect of the system's uncertainties.


Next, we evaluated the reliable logic computation probability, considering a fixed propagation delay of $100$ seconds and different molecular input signals concentration $m_B$ and $m_C$ (ranging from $1\mu$mol/L to $3\mu$mol/L), which is shown in Figure \ref{fig:theo_results} (b). In this scenario, the ON-OFF switch dramatically decrease its performance if more molecular signal is input into the system, with the no production delay cases performing better than the cases where the production delay was considered. Therefore, for minimal effects on the overall performance of this gate, only a small range of molecular input signal can be considered (from $1.12$ mmol/L to $1.60$ mmol/L). Figure \ref{fig:theo_results} (b) also shows that the AND gate reliable logic computation probability reaches a plateau for molecular input signal concentrations above $1.75$ mmol/L for the no production delay case and $2$ mmol/L when there is delay on the production of the molecular output signal. These results shows that the logic computation using AND gates is more robust against a wide range of molecular input signal concentrations, but it can suffer considerably for lower difference between the production and propagation delays. On the other hand, the ON-OFF switch tends to be more robust against the delay difference and for a small range of molecular input signal concentrations. 
 

\subsubsection{Blind Detector Threshold Analysis}\label{sec:blind_thold_analysis}

We also evaluate the reliable logic computation probability considering the blind detector threshold defined in (\ref{eq:tholdblind_init})-(\ref{eq:tholdblind_end}) and the same scenarios of the standard detector threshold (see Section \ref{sec:std_thold_analysis}). Figure \ref{fig:theo_results} (c) shows the scenario where different propagation delay values are considered. For both gates, AND gate does not have its reliable logic computation probability affected by the propagation delay. This result is due to the different detection process, which is more robust to the possible variations caused by the propagation delay. Despite that, this same technique is not robust enough for the ON-OFF switch, which improves its reliability when considering higher propagation delays. When comparing the reliable logic computation probability of the ON-OFF switch depicted in Figures \ref{fig:theo_results} (a) and \ref{fig:theo_results} (c), one can note that they are similar, which shows this specific gate can use both detector thresholds without affecting the BMCoC's performance.

Following the propagation delay analysis, we investigate the impact of the different molecular input signal concentrations $m_B$ and $m_C$ (the same values applied for the standard detector threshold analysis) on the reliability of logic computation of both gates, when considering the blind detector threshold, see Figure \ref{fig:theo_results} (d). In this scenario, the AND gate shows a different performance when compared with the standard detector threshold. The molecular input signal concentrations considered in this analysis are in the plateau range of the AND gate, which happen for values above $1.75$ mmol/L for Figure \ref{fig:theo_results} (b). On the other hand, the ON-OFF switch has high reliability for a small range of molecular input signal concentrations with a range of $1.12$ mmol/L to $1.60$ mmol/L. Therefore, the ON-OFF switch performance shows the same behaviour for both scenarios when comparing Figures \ref{fig:theo_results} (b) and \ref{fig:theo_results} (d). These results showed that the blind detector threshold gave stability for the digitalisation process of the molecular output signal produced by the AND gate and did not affect the performance of the molecular output signal detection originated from the ON-OFF switch. 

 \begin{figure}[t!]
 \centering
 \includegraphics[width=0.8\columnwidth]{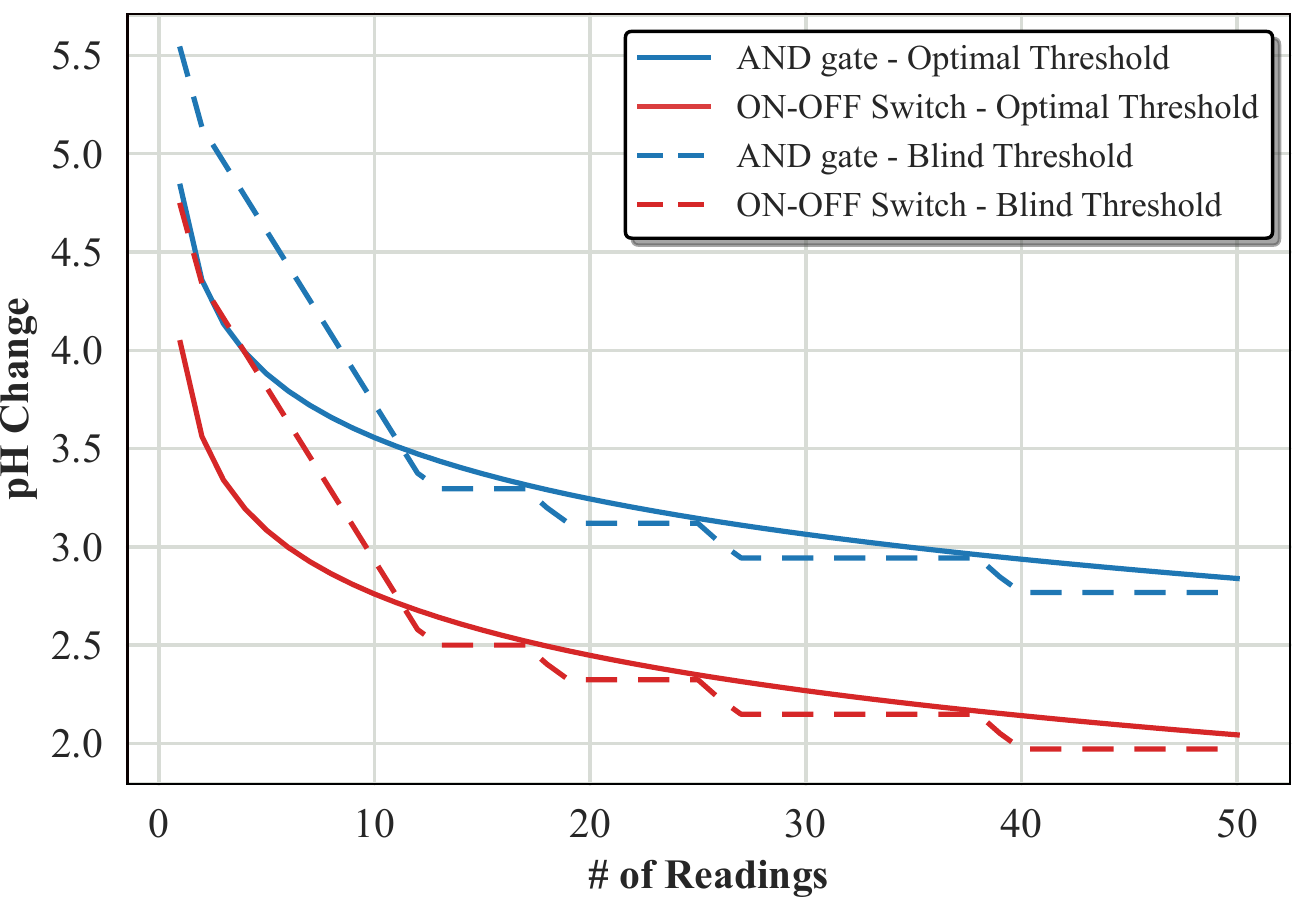}
 \caption{Evaluation of the pH change saturation due to the cumulative addition of the molecular signal produced by the engineered bacteria on the fluid media around the electrochemical sensor for the AND gate and ON-OFF switch after multiple readings, for both proposed detection thresholds.}
 \label{fig:saturation}
 \end{figure}

\subsubsection{Saturation}

 The number of readings that the sensor can perform before losing its sensitivity is an essential parameter for the proposed BMCoC system. Therefore, based on the results obtained by the electrochemical sensor detection thresholds analysis, we can evaluate this limit for the pH variation detection. For example, by adding $2.27\times10^{-8}$ protons, we would change the pH of fluid from $9$ to $7.62$. If repeating this process one more time, the pH would reduce from $7.62$ to $7.33$.  By repeating this process several times, the pH reduction will become so small that it becomes hard to be detected by the electrochemical sensor. Therefore, we evaluate this saturation of the electrochemical sensor for the cumulative molecular output signal concentration as shown in Figure \ref{fig:saturation}. Both the standard and blind detector thresholds shows an exponential-like decrease for the pH variation, as this saturation model follows the same calculations from (\ref{eq:pH7.62})-(\ref{eq:pH6.91}). However, the standard detector threshold has a smoother curve than the blind detector threshold. Moreover, despite showing a more abrupt pH reduction for the initial 12 readings, the blind threshold showed a similar pH reduction for the remaining measurements. This result suggests that even electrochemical sensors with lower sensitivity (only detect high pH variations) can use the blind detector threshold.

\section{Conclusion}\label{sec:conclusions}

In this paper, we introduced the experimental and theoretical analysis of a Bacterial Molecular Computing on a Chip that contains engineered cells performing computation based on input molecular signals. 
The molecular signal output from the engineered bacterial population modifies the pH or the dissolved oxygen concentration, which is sensed by the electrochemical sensors used for detection. The paper also investigated the computation reliability of the BMCoC under the impact of unwanted effects, i.e., molecular production and propagation delays, as well as electrolyte noise. Our experimental analysis showed the engineered bacterial population's ability to detect molecular signals (e.g., acetate and propionate) at low concentrations and can produce required output to be detected by the electrochemical sensors. 
Furthermore, the electrochemical sensors showed a linear dependency on the pH and dissolved oxygen concentration variation, granting certain robustness to these sensors. 
We also found that considering a detection threshold of $2.27\times10^{-8}$ mol/L (or pH $7.64$), the electrochemical sensors can operate better if the fluid medium has a pH 9 instead of a pH 7. This threshold result into a minimal electrical current level ($I_c=0.45\,\text{nA}$) for the molecular output signal concentration detection. We also show that the AND gate have a more robust logic computation performance than the ON-OFF switch for the analysed scenarios. However, the ON-OFF switch could achieve a higher reliable logic computation probability than the AND gate for these same scenarios. The BMCoC can open up to a plethora of different diagnostic applications, and can lay the foundation for the development of future Internet of Bio-Nano Things.


\ifCLASSOPTIONcompsoc
  \section*{Acknowledgments}
\else
  \section*{Acknowledgment}
\fi

This work was funded by Science Foundation Ireland and the Department of Agriculture, Food, and Marine via the VistaMilk research centre (grant no. 16/RC/3835).

\appendix
\subsection{Engineered Bacteria}

We designed a wet lab experiment where a AND gate process nitrile and IPTG molecular concentrations to output ammonia and hydrogen, that would change the pH of the media around the electrochemical sensors. The detection of AND logic operation can be done by measuring the ammonia (spectroscopy) or the hydrogen ions (pH variation). In this experiment, we investigate ammonia production by using Nessler’s microscale ammonia assay \cite{Coady2013}. Assays were carried out in $150\,\mu\text{L}$ format containing potassium phosphate buffer pH 7, a final cell O.D. $\text{@}600\,\text{nm}$ of $0.5$ and a final substrate (nitrile) concentration of $0\,\text{mmol/L}$, $5\,\text{mmol/L}$ or $10\,\text{mmol/L}$ (no, low or high molecular output signal amplitude, respectively). The amounts of enzyme expression inducer, IPTG (Isopropyl-$\beta$-D-thiogalactopyranoside, Zymo Research, L1001-5), was either $0\,\text{mmol/L}$, $0.1\,\text{mmol/L}$ or $0.5\,\text{mmol/L}$ (no, low or high IPTG, respectively) to test the operation of the AND logic gate. The reaction was carried out over $5$ hours with regular readings taken throughout. To quench the continued generation of signal, $37.5\,\mu\text{L}$ of $250\,\text{mmol/L}$ HCl (Sigma-Aldrich, Cat. No. 435570) was added to stop the reaction. Cell biomass was removed at $500$ x g for $10$ minutes at $4\,^{\circ}\text{C}$ to be able to quantify the molecular output concentration. We transferred $20\,\mu\text{L}$ of the quenched reaction supernatant to a microtiter plate and to this $181\,\mu\text{L}$ of the Nessler’s master mix was added ($151\,\mu\text{L}$ deionised $H_2O$, $1.0\,\mu\text{L}$ 10N NaOH (Sigma-Aldrich, Cat. No. 765429) and $25\,\mu\text{L}$ Nessler’s reagent (Sigma-Aldrich, 72190). The reaction supernatant was incubated at room temperature ($22\,^{\circ}\text{C}$) for $10$ minutes and the absorbance was read at $425\,\text{nm}$. 

A nitrilase gene from a \emph{Burkholderia} bacteria was PCR amplified and cloned to an expression vector to process molecules as an AND logic gate.  Each $15\,\mu\text{L}$ PCR reaction mixture contained $7.5\,\mu\text{L}$ Platinum\texttrademark SuperFi\texttrademark Green PCR Master Mix (ThermoFisher Scientific, Cat. No. 12359010), $15\,\mu\text{M}$ of each primer and $1\,\mu\text{L}$ cell suspension with a final cell O.D. $\text{@}600\,\text{nm}= 0.04$. The following PCR conditions were used: 1 cycle of $95\,^{\circ}\text{C}$ for $5$ min, $30$ cycles of $95\,^{\circ}\text{C}$ for $1$ min, $56\,^{\circ}\text{C}$ for $1$ min, $72\,^{\circ}\text{C}$ for $2$ min, followed by $1$ cycle of $72\,^{\circ}\text{C}$ for $5$ min. The PCR product was cleaned using the Zymo Research clean and concentrator\texttrademark -5 (Zymo Research, Cat. No. D4013) as per the manufacturer instructions with elution in water. The pRSF-2 Ek/LIC vector (Novagen, Cat. No. 71364) was used for expression of the nitrilase. Cloning procedures were followed as per manufacturer’s instructions, with ligations transformed to E. coli BL21 (DE3) as per manufacturer’s guidelines for heat shock transformations. 

For the ON-OFF switch, the bacterial population was created by cultivating, extracting and immobilizing acetate and propionate grown A11 and P1 cells, respectively. This engineering process is similar to the methods previously described by \cite{Sweeney2018}, however, the following modifications were made for this study: 1.) IMD Wldgyep (acetate biosensor strain) and IMD Wldgyepak (propionate biosensor) strains were replaced with A11 and P1 \emph{E. coli} strains, respectively. A11 and PI cells differ in that their ability to catabolise formate has also been removed; 2.) The previously used dissolved oxygen (DO) probe was updated with an array of Vernier (DO-BTA) DO probes. These probes were selected due to their affordability, their increased sampling frequency ($1$ Hz) and their ability to be interfaced as an array using Arduino microcontrollers. For Vernier DO probes to be made compatible with the previously developed cell immobilization technique, Vernier probe tips needed to be modified slightly. Vernier DO probe tips are concave in shape which results in their oxygen permeable membranes being exposed. To project the membrane, Vernier probe tips possess raised bevel rings. Immobilized cells can only be fixed to a smooth probe tip surface and as such Vernier’s protective bevels had to be removed by a scalpel. To ensure that fixed cells are protected from magnetic stirrer flea impacts in the absence of raised levels, perforated steel guards were fabricated and fitted on top of the cells.

\subsection{Electrochemical Sensors}
The pH change will be detected by the electrochemical sensors measuring the oxidation/reduction potentials of a signalling mediator molecule. A similar process was applied by Wahl et. al to simulate the electric current on a gold nanowire electrode \cite{Wahl2014}. Furthermore, we use \emph{ferrocene monocarboxylic acid} (FcCOOH), a signalling molecule, to detect the production of protons (in this case, $H^+$) and the consequent decrease in pH \cite{Wahl2014,Raoof2007}. The redox potential $E_0$ of FcOOH has a pH dependence of $+30$mV/pH at a polypyrrole modified reference electrode \cite{Wahl2014}.  FcCOOH at a polypyrrole reference electrode has a higher redox potential for high pH levels, meaning that when more protons are present, a smaller electric potential is required for oxidisation.

Electrochemical oxygen reduction was carried out in water. The water sample was allowed to saturate with oxygen through exposure to air at room temperature ($17\,^{\circ}\text{C}$). Low concentration oxygen samples were prepared by purging the water solutions with nitrogen for $30$ minutes. After purging, the water samples were covered with parafilm which contained a small perforation. This allowed oxygen to dissolve back into the solution, albeit at a slow rate. It was found that oxygen typically re-dissolved at the rate of approximately $1$ ppm per $3$ minutes when the concentration was between $0$ and $4$ ppm. After $4$ ppm was reached, the rate slowed down to $1$ ppm per $7$ minutes until a $6$-ppm concentration was reached. To bring the solution back to saturated concentrations, oxygen had to be bubbled into the solution, as the time to reach saturation by diffusion was considerably longer. This resulted in water solutions with various oxygen concentration between fully purged ($0.5$ ppm), and fully saturated ($8.8$ ppm). All oxygen concentrations were measured using a commercial optical DO probe (Hach, LDO101). The electrochemical analysis was carried out using an Autolab potentiostat, with the electrochemical cell kept in a Faraday cage. A gold-gold IDE array was used for this work. 

\ifCLASSOPTIONcaptionsoff
  \newpage
\fi

\bibliographystyle{IEEEtran}
\bibliography{TBioCas_Computing}
\end{document}